\documentclass[aps,prc,showpacs,twocolumn]{revtex4}
\usepackage{graphicx}
\begin{document}

\def\pt{$p_T$}
\def\aa{$AuAu$ }
\def\pa{$pA$ }
\def\da{$dAu$ }
\def\sji{$S^j_i$ }
\def\q2{$Q^2$ }
\def\gev{GeV}
\def\kns{$K_{NS}$ }

\title{Recombination of Shower Partons  in
Fragmentation Processes}
\author{Rudolph C. Hwa$^1$ and  C.\ B.\ Yang$^{1,2}$}
\affiliation{$^1$Institute of Theoretical Science and Department of
Physics, University of Oregon, Eugene, OR 97403-5203, USA}
\affiliation{
$^2$Institute of Particle Physics, Hua-Zhong Normal University,
Wuhan 430079, P.\ R.\ China}
\date{}
\begin{abstract}
We develop the notion of shower partons and determine their
distributions in jets in the framework of the recombination model.
The shower parton distributions obtained render a good fit of the
fragmentation functions. We then illustrate the usefulness of the
distributions in a problem where a jet is produced in the environment
of thermal partons as in heavy-ion collisions. The recombination of
shower and thermal partons is shown to be more important than
fragmentation. Application to the study of two-particle correlation
in a jet is also carried out.

\end{abstract}
\maketitle

\section{Introduction}
The theoretical description of hadron production at large transverse
momentum ($p_T$) in either hadronic or nuclear collisions at high
energies is traditionally framed in a two-step process that involves first
a hard scattering of partons, followed by the fragmentation of the
scattered parton to the detected hadron \cite{rdf,sw}.  The first part is calculable in
perturbation QCD, while the second part makes use of fragmentation
functions that are determined phenomenologically.  Such a production
mechanism has recently been found to be inadequate for the production
of particles at intermediate $p_T$ in heavy-ion collisions \cite{hy,gr,fr}.
Instead of fragmentation it is the recombination of partons that is
shown to be the more appropriate hadronization process, especially
when the soft partons are involved.  Although at extremely high
$p_T$ fragmentation is still dominant, it is desirable to have a universal
description that can be applied to any $p_T$, based on the same
hadronization scheme.  To achieve that goal it is necessary that the
fragmentation process can be treated as the result of recombination of
shower partons in a jet.    The purpose of this paper is to take that first
step, namely:  to introduce the notion of shower partons and to
determine their distributions in order to represent the
phenomenological fragmentation functions in terms of recombination.

The subject matter of this work is primarily of interest only to
high-energy nuclear collisions because hadronization in such processes
is always in the environment of soft partons.  Semi-hard shower partons
initiated by a hard parton can either recombine among themselves or
recombine with soft partons in the environment.  In the former case the
fragmentation function is reproduced, and nothing new is achieved.  It
is in the latter case that a very new component emerges in heavy-ion
collisions, one that has escaped theoretical attention thus far.  It should
be an important hadronization process in the intermediate $p_T$
region.  Our main objective here is to quantify the properties of shower
partons and to illustrate the importance of their recombination with
thermal partons.  The actual application of the shower parton
distributions (SPD) developed here to heavy-ion collisions will be
considered elsewhere \cite{hy2}.

The concept of shower partrons is not new, since attempts have been made to
generate such partons in pQCD processes as far as is permitted by the
validity of the procedure. Two notable examples of such attempts are the
work of Marchesini and Webber \cite{mw} and Geiger \cite{ge}. However,
since pQCD cannot be used down to the hadronization scale, the branching or
cascading processes terminate at the formation of color-singlet pre-hadronic
clusters, which cannot easily be related to our shower partons and their
hadronization. We shall discuss in more detail at the end of Secs.\ III and
IV the similarities and differences in the various approaches.

\section{Recombination Model for Fragmentation}

The fragmentation of a parton to a hadron is not a process that can be
calculated in pQCD, although the $Q^2$  evolution of the fragmentation
function (FF) is calculable.  The FF's are usually parameterized by fitting
the data from $e^+e^-$ annihilations \cite{bkk,kkp,kkp2} as well as
from $p\bar{p}$ and $e^\pm p$ collisions \cite{kkp2}.  Although the
QCD processes of generating a parton shower by gluon radiation and
pair creation cannot be tracked by perturbative methods down to low
virtuality, we can determine the SPD's phenomenologically in much the
same way that the FF's themselves are, except that we fit the FF's,
whereas the FF's are determined by fitting the data.  An important
difference is that both the shower partons and their distributions are
defined in the context of the recombination model, which is the key link
between the shower partons (inside the black box called FF) and the
observed hadron (outside the black box).

In the recombination model the generic formula for a hadronization
process is \cite{rh}
\begin{eqnarray}
xD(x) = \int^x_0 {dx_1 \over x_1} \int^x_0 {dx_2 \over
x_2}F_{q\bar{q}'} (x_1, x_2) R (x_1, x_2, x) \ ,
\label{1}
\end{eqnarray}
where $F_{q\bar{q}'} (x_1, x_2)$ is the joint distribution of a quark
$q$ at momentum fraction $x_1$ and an antiquark $\bar{q}'$ at
$x_2$, and $R (x_1, x_2, x)$ is the recombination function (RF) for the
formation of a meson at $x$.  We have written the LHS of Eq.\ (\ref{1})
as $xD(x)$, the invariant FF, but the RHS would have the same form if
the equation were written for the inclusive distribution, $xdN/dx$, of a
meson produced in a collisional process.   In the former case of
fragmentation, $F_{q\bar{q}'}$ refers to the shower partons initiated by
a hard parton.  In the latter case of inclusive production,
$F_{q\bar{q}'}$ refers to the $q$ and $\bar{q}'$ that are produced by
the collision and are to recombine in forming the detected meson.  The
equations for the two cases are similar because the physics of
recombination is the same.  In either case the major task is in the
determination of the distribution $F_{q\bar{q}'}$.

We now focus on the fragmentation problem and regard Eq.\ (\ref{1})
as the basis of the recombination model for fragmentation.  The LHS is
the FF, known from the parameterization that fits the data.  The RHS
has the RF that is known from previous studies of the recombination
model \cite{rh,hy3} and will be specified in the next section.  Thus it is
possible to determine the properties of $F_{q\bar{q}'}$ from Eq.\
(\ref{1}).  To facilitate that determination we shall assume that
$F_{q\bar{q}'}$ is factorizable except for kinematic constraints, i.e.,
in schematic form we write it as
\begin{eqnarray}
F^{(i)}_{q\bar{q}'}(x_1, x_2) = S^q_i(x_1) S_i^{\bar{q}'}(x_2) \quad ,
\label{2}
\end{eqnarray}
where $S^q_i(x_1)$ denotes the distribution of shower parton $q$ with
momentum fraction $x_1$ in a shower initiated by a hard parton $i$.
The exact form with proper kinematic constraints will be described in
detail in the next section.  Here we remark on the general implications
of Eqs.\ (\ref{1}) and (\ref{2}).

The important point to emphasize is that we are introducing the notion
of shower partons and their momentum distributions $S^j_i(x_1)$.  The
significance of the SPD is not to be found in problems that involve only
the collisions of leptons and hadrons, for which the fragmentation of
partons is known to be an adequate approach, and the recombination of
shower partons merely reproduces what is already known.  The
knowledge about the SPD becomes crucial when the shower partons
recombine with other partons that are not in the jet but are in the
ambient environment.  We shall illustrate this important point later.

It should be recognized that the SPD that we shall determine through
the use of Eqs.\ (\ref{1}) and (\ref{2}) depends on the specific form of
$R (x_1, x_2, x)$, which in turn depends on the wave function of the
meson produced.  It would be inconsistent to use our $S^j_i$ given
below in conjunction with some approximation of the RF that differs
significantly from our $R$.  The recombination of two shower partons
must recover the FF from which the SPD's are obtained.

Finally, we remark that $S^j_i$ should in principle depend on $Q^2$  at
which the $D(x, Q^2)$ is used for its determination, since $Q^2$
evolution affects both.  It is outside the scope of this paper to treat the
$Q^2$  dependence of $S^j_i$.  Our aim here is to show how $S^j_i$ can
be determined phenomenologically, and how it can be applied, when
$Q^2$  is fixed.  The same method can be used to determine $S^j_i$ at
other values of $Q^2$ .  In practice, the $Q^2$  dependence of
$S^j_i$ is not as
important as the inclusion of the role of the shower partons in the
first place at any reasonably approximate $Q^2$  in heavy-ion collisions
where hard partons are produced in a  range of transverse
momentum.

\section{Shower Parton Distributions}

In order to solve Eqs.\ (\ref{1}) and (\ref{2})  for $S^j_i$, we first point
out that there are various $D(x)$ functions corresponding to various
fragmentation processes.  We shall select five of them, from which we
can determine five SPD's.  Three of them form a closed set that involves
no strange quarks or mesons.  Let us start with those three.  Consider
the light quarks $u$, $d$, $\bar{u}$, $\bar{d}$, and gluon $g$.  They
can all fragment into pions.  To reduce them to three essential FF's, we
consider the three basic types $D^{\pi}_V$, $D^{\pi}_S$
and $D^{\pi}_G$, that correspond to  valence, sea and gluon
fragmentation, respectively.  If the fragmenting quark has the same
flavor as that of a valence quark in $\pi$, then the valence part of the
fragmentation is described by $D^{\pi}_V$, e.g., $u_v\rightarrow \pi
^+$, $d_v \rightarrow \pi ^-$, $\bar{u}_v \rightarrow \pi^-$, the
subscript $v$ referring to the valence component.  All other cases of
quark fragmentation are described by  $D^{\pi}_S$, e.g.,
$u\rightarrow \pi ^-$, $d\rightarrow \pi ^+$, $\bar{u}
\rightarrow \pi^+$.  If the initiating parton is a gluon, then we have
$D^{\pi}_G$ for any state of $\pi$.  Those FF's are given by Ref.\
\cite{bkk} in parametric form.  We shall use them even though they are
older than the more recent ones \cite{kkp,kkp2,sk}, which do not give
the $D^{\pi}_V$ and $D^{\pi}_S$ explicitly.  Our emphasis here is not
on accuracy, but on the feasibility of extracting the SPD's from the
FF's of the type discussed above.   We shall determine $S^j_i$
from the BKK parameterization \cite{bkk} with $Q^2$  fixed at 100
GeV$^2$ and  demonstrate  that the use of shower partons is
important
 in heavy-ion collisions.

For the SPD's we use the notation $K_{NS}$ and $L$ for valence and
sea-quark distributions, respectively, in a shower initiated by a
quark or antiquark,
  and $G$ for any light quark distribution in a
gluon-initiated
shower.  That is, for example, $K_{NS}$ $= S^{u_v}_u$, $L = S^d_u$, $G
= S^u_g$.  It should be recognized that $L$ also describes the sea quarks
of the same flavor, such as $S^{u_{sea}}_u$, so that the overall
distribution of shower quark that has the same flavor as the initiating
quark (e.g. $u
\rightarrow u$) is given by
\begin{eqnarray}
K = K_{NS} + L \quad .
\label{3}
\end{eqnarray}

  It is evident from the above discussion that there is a closed
relationship that is independent of other unknowns.  It follows from Eq.
(\ref{1}) when restricted to sea-quark fragmentation:
\begin{eqnarray}
xD^{\pi}_S(x) = \int {dx_1 \over x_1} {dx_2 \over
x_2}L(x_1) L \left({x_2\over 1 - x_1}\right) R_{\pi} (x_1, x_2, x) \quad .
\label{4}
\end{eqnarray}
The sea-SPD $L(z)$ can be determined from this equation alone.  In Eq.\
(\ref{4}) we have exhibited the argument of the second $L$ function
that reflects the momentum constraint, i.e., if one shower parton has
momentum fraction $x_1$, then the momentum fraction of the other
recombining shower parton cannot exceed $1 - x_1$, and can only be a
fraction of the balance $x_2/(1 - x_1)$.   Symmetrization of $x_1$ and
$x_2$ is automatic by virtue of the invariance of $R_{\pi} (x_1, x_2,
x)$ under the exchange of $x_1$ and $x_2$.

After $L(z)$ is determined from Eq.\ (\ref{4}), we next can obtain
$K_{NS}$ from
\begin{eqnarray}
xD^{\pi}_V(x) = \int {dx_1 \over x_1} {dx_2 \over
x_2} \left\{K_{NS} (x_1), L(x_2) \right\} R_{\pi}
(x_1, x_2, x) ,
\label{5}
\end{eqnarray}
where the curly brackets define the symmetrization of the leading
parton momentum
\begin{widetext}
\begin{eqnarray}
\left\{K_{NS} (x_1), L(x_2) \right\} \equiv {1 \over 2} \left[K_{NS}
(x_1) L\left({ x_2\over 1-x_1}\right) + K_{NS}\left({ x_1\over
1-x_2}\right)  L (x_2)\right] \quad .
\label{6}
\end{eqnarray}
\end{widetext}
Finally, we have the closed equation for the gluon-initiated shower
\begin{eqnarray}
xD^{\pi}_G(x) = \int {dx_1 \over x_1} {dx_2 \over
x_2}G (x_1) G\left({ x_2\over 1-x_1}\right) R_{\pi} (x_1, x_2, x) .
\label{7}
\end{eqnarray}
In this non-strange sector we have 3 SPD's ($L$, $K_{NS}$ and $G$) to
be determined from the 3 phenomenological FF's ($D^{\pi}_S$,
$D^{\pi}_V$ and $D^{\pi}_G $).

In extending the consideration to the strange sector, we must make use
of $L$ and $G$ determined in the above set and two new FF's $D^K_S$
and $D^K_G$ that describe the fragmentation of a non-strange quark
and gluon, respectively, to a kaon.  That is, we have
\begin{eqnarray}
xD^K_S(x) = \int {dx_1 \over x_1} {dx_2 \over
x_2} \left\{L (x_1), L_s(x_2)\right\}R_K
(x_1, x_2, x) ,
\label{8}
\end{eqnarray}
\begin{eqnarray}
xD^K_G(x) = \int {dx_1 \over x_1} {dx_2 \over
x_2} \left\{G (x_1), G_s(x_2)\right\}R_K
(x_1, x_2, x) ,
\label{9}
\end{eqnarray}
where $L_s$ and $G_s$ are two additional SPD's specifying the strange
quark distributions in showers initiated by non-strange and gluon
partons, respectively.  $R_K$ is the RF for kaon.

To complete the description of the integral equations, we now specify
the RF's.  They depend on the square of the wave functions of the
mesons, $\pi$ and $K$, whose structures in momentum space have
been quantified in the valon model \cite{rh,hy3}.  Unlike the case of the
proton, whose structure is well studied by deep inelastic scattering so
that the valon distribution can be obtained from the parton distribution
functions \cite{hy4}, the RF for the pion relies on the parton
distribution of the pion probed by Drell-Yan process \cite{sut}.  The
derivation of the RF's for both $\pi$ and $K$ is given in \cite{hy3};
they are
\begin{eqnarray}
R_{\pi} (x_1, x_2, x) = {x_1 x_2  \over  x^2} \delta \left({x_1
\over  x} + {x_2  \over x } - 1\right),
\label{10}
\end{eqnarray}
\begin{eqnarray}
R_K (x_1, x_2, x) = 12 \left({x_1 \over  x} \right)^2 \left({x_2
\over  x} \right)^3 \delta \left({x_1
\over  x} + {x_2  \over x } - 1\right),
\label{11}
\end{eqnarray}
The $\delta$ functions guarantee the momentum conservation of the
recombining quarks and antiquarks, which are dressed and become the
valons of the produced hadrons.

Since the recombination process involves the quarks and antiquarks,
one may question the fate of the gluons.  This problem has been treated
in the formulation of the recombination model \cite{rh}, where gluons
are converted to quark-antiquark  pairs in the sea before
hadronization.  That is,
the sea is saturated by the conversion to carry all the momentum, save
the valence parton.  Such a procedure has been shown to give the correct
normalization of the inclusive cross section of hadronic collisions
\cite{rh}.  In the present problem of parton fragmentation we
implement the recombination process in the same framework, although
gluon conversion is done only implicitly.  What is explicit is that the gluon
degree of freedom is not included in the list of shower partons.  It
means that in the equations for $D^{\pi}_V$, $D^{\pi}_S$,
   and $D^{\pi}_G$ (and likewise in the strange sector) only $K_{NS}$,
$L$ and $G$ appear; they are the SPD's of quarks and antiquarks that
are to recombine.  Those quarks and antiquarks must include the
converted sea, since they are responsible for reproducing the FF's
through Eqs.\ (\ref{4}), (\ref{5}) and (\ref{7}) without gluons.  Thus the
shower partons whose momentum distributions we calculate are defined
by those equations that have no gluon component for recombination,
and would not be the same as what one would conceptually get (if it
were possible)
in a pQCD calculation that inevitably has both quarks
and gluons.

It should be noted that our procedure of converting gluons to $q\bar q$
pairs is essentially the same as what is done in \cite{mw}, whose branching
processes terminate at the threshold of the non-perturbative regime. In
that approach nearby quarks and antiquarks that are the products of the
conversion from different gluons form color-singlet clusters of various
invariant masses that subsequently decay (or fragment as in strings)
sequentially through resonances to the lowest lying hadron states
\cite{bw}. Similar but not identical approach is taken in \cite{eg}, where
gluons are not directly converted to $q\bar q$ pairs, but are either
absorbed or annihilated by $g+g\rightarrow q+\bar q$ Born-diagram processes.

\section{Results}

We now proceed to solve the integral equations for the five FF's, which
are known from Ref.\ \cite{bkk}.  Those equations relate them to the
five unknown SPD's:  $K_{NS}$, $L$, $G$, $L_s$ and $G_s$.
If those equations were algebraic, we obviously could solve them for the
unknowns.  Being integral equations, they can nevertheless be ``solved''
by a fitting procedure that should not be regarded as being
unsatisfactory for lack of mathematical rigor, since the FF's themselves
are obtained by fitting the experimental data in some similar manner.
Indeed, the FF's in the next-to-leading order are given in parameterized
forms \cite{bkk}
\begin{eqnarray}
D^h_k (x) = N x^{\alpha}(1 - x)^{\beta}(1 +
x)^{\gamma}
\label{12}
\end{eqnarray}
where the parameters for $Q^2=100$ GeV$^2$ are
given in Table I for $k = S, V, G$ and $h = \pi, K$.

\begin{table}[tbph]
\caption{Parameters in Eq.\ (\ref{12}) for $Q^2=100 $GeV$^2$.}
\begin{center}
\begin{tabular}{|l||c|c|c|c|} \hline
      &$N$&$\alpha$&$\beta$&$\gamma$\\ \hline\hline
$D^{\pi}_S$&2.7236&$-0.734$&3.384&$-5.471$\\ \hline
$D^{\pi}_V$&0.2898&$-1.040$&1.608&$-0.111$\\ \hline
$D^{\pi}_G$&0.7345&$-1.112$&2.547&$-0.541$\\ \hline
$D^K_S$&0.2106&$-1.005$&2.548&$-0.620$\\ \hline
$D^K_G$&0.0768&$-1.481$&2.489&$-0.778$\\ \hline
\end{tabular}
\end{center}
\end{table}

\begin{figure}[tbph]
\includegraphics[width=0.45\textwidth]{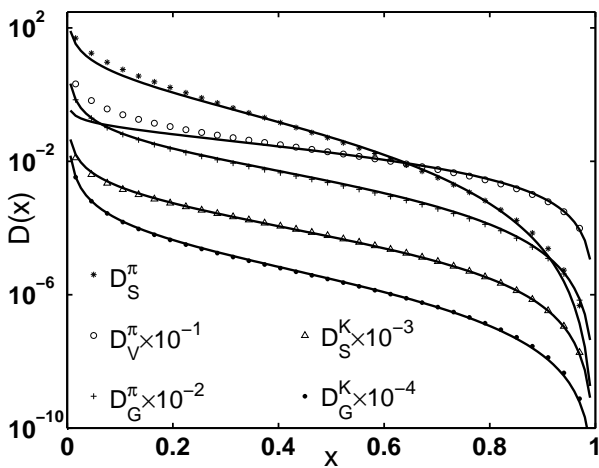}
\caption{
Fragmentation functions, as parametrized in
\cite{bkk}, are shown in symbols, while those calculated in the
recombination model are shown by the solid lines.  All curves are
for $Q^2 = 100$GeV$^2$.}
\end{figure}

All five SPD's are denoted  collectively by  $S^j_i$ with $i = q,
\bar{q}, g$ and $j = q, s, \bar{q}, \bar{s}$, where $q$ can be either
$u$ or $d$.  If in $i$ the initiating hard parton is an $s$ quark, it is
treated as $q$.  That is not the case if $s$ is in the produced shower.
Our parametrization of $S_i^j$ has the form
\begin{eqnarray}
S_i^j(z) = Az^a (1-z)^b (1 + cz^d),
\label{13}
\end{eqnarray}
where the dependences of the parameters $A$, $a$,
etc. on $i$ and $j$ are not exhibited explicitly, just as in Eq.\
(\ref{12}). Substituting Eqs.\ (\ref{10}) and (\ref{11}) into
(\ref{4})-(\ref{9}), we can determine the parameters one equation
at a time, i.e., $L$ from
$D^{\pi}_S$, $G$ from $D^{\pi}_G$, and then $K_{NS}$ from
$D^{\pi}_V$ and so on.  In most cases it can be shown that the $x
\to 1$ limit requires $b = \beta$.  The final results of the fits
are shown in Fig.\ 1 with the corresponding parameters given in
Table II.

\begin{table}[tbph]
\caption{Parameters in Eq.\ (\ref{13}).}
\begin{center}
\begin{tabular}{|l||c|c|c|c|c|} \hline
      &$A$&$a$&$b$&$c$&$d$\\ \hline\hline
$K_{NS}$&0.333&0.45&2.1&5.0&0.5\\ \hline
$L$&1.881&0.133&3.384&$-0.991$&0.31\\ \hline
$G$&0.811&$-0.056$&2.547&$-0.176$&1.2\\ \hline
$L_s$&0.118&$-0.138$&2.3&0.90&0.1\\ \hline
$G_s$&0.069&$-0.425$&2.489&$-0.5$&1.1\\ \hline
\end{tabular}
\end{center}
\end{table}

\begin{figure}[tbph]
\includegraphics[width=0.45\textwidth]{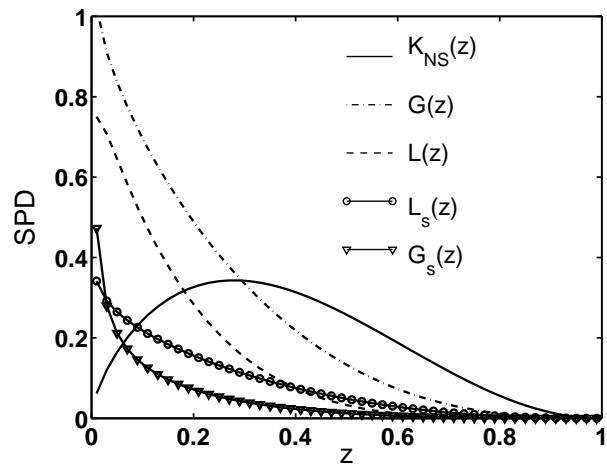}
\caption{
Shower parton distributions determined in the
recombination model, corresponding to the parameterization
given in Table II.}
\end{figure}

It is evident from Fig.\ 1 that all the fits are very good, except
in the low $x$ region of $D^{\pi}_V(x)$.  In the latter case we are
constrained by the condition
\begin{eqnarray}
\int {dz \over z} K_{NS}(z) = 1
\label{14}
\end{eqnarray}
that is imposed by the requirement that there can be
only one valence quark in the shower partons.  However, the fit for
$x > 0.4$ is excellent, and that is the important region for the
determination of
$K_{NS}(z)$.  In application to $u \to \pi ^+$, say, the $u$ quark
in the shower must have both valence and sea quarks so the shower
distribution for the
$u$ quark is always the sum:
$K(z) = K_{NS}(z) + L (z)$.  Since $L(z)$ is large at small $z$,
and is accurately determined, the net result for $K(z)$ should be
quite satisfactory.

It is remarkable how well the FF's in Fig.\ 1 are reproduced in the
recombination model.  The corresponding SPD's that make possible
the good fit are shown in Fig.\ 2.  They have very reasonable
properties, namely:  (a) valence quark is harder (b) sea quarks are
softer, (c) gluon jet has higher density of shower partons, and (d)
the density of produced $s$ quarks is lower than that of the light
quarks.

\begin{figure}[tbph]
\includegraphics[width=0.45\textwidth]{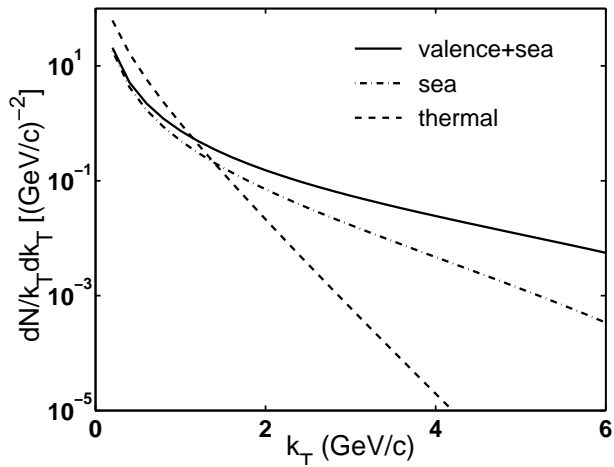}
\caption{
Parton distributions in transverse momentum $k_T$ for
valence+sea quark (solid line), sea quark (dash-dot line) and thermal
partons (dashed line).}
\end{figure}

It is appropriate at this point to relate our approach to those of Marchesini-Webber
\cite{mw} and Geiger \cite{ge}, which are serious attempts to incorporate
the QCD dynamics in their description of the branching and collision
processes. The former is done in the momentum space only, whereas the
latter is formulated in space-time as well as in momentum space. The parton cascade
model of Geiger is a very ambitious program that treats a large variety of
processes ranging from $e^+e^-$ annihilation \cite{eg} to deep inelastic
scattering \cite{egk} to hadronic and nuclear collisions \cite{kgs}. The
evolution of partons is tracked by use of relativistic transport equations with gain and
loss terms. Cluster formation takes into account the invariant distance
between near-neighbor partons. Cluster decay makes use of the Hagedorn
spectrum and the particle data table. Because of the complexity of the
problems both QCD models are implemented by Monte Carlo codes. The
predictive power of the models is exhibited as numerical outputs that
cannot easily be adapted for comparison with our results on the SPD's. Our
approach makes no attempt to treat the QCD dynamics; however, the SPD's
obtained are guaranteed to reproduce the FF's on the one hand, and are
conveniently parameterized for use in other context that goes beyond
fragmentation, as we shall show in the next section. From the way the
color-singlet clusters are treated in the QCD models, it is clear that our
shower partons do not correspond to the partons of those models at the end of their evolution processes, except in the special case when the cluster consists of only one particle. In our approach the non-perturbative part of how the shower partons dress themselves and recombine to form hadrons with the proper momentum-fraction distributions is contained in the RF's. Such shower partons that are ready to hadronize are sufficiently far from other shower partons as to be independent from them. In general, they cannot be identified with the $q$ and $\bar q$ that form the color-singlet clusters in the QCD models, but are more closely related to the constituents of the final hadrons, as in the case of quarkonium formation \cite{kge}. The distribution of those constituents in a hard-parton shower cannot be displayed in the QCD models, but are determined by us by solving Eqs.\ (\ref{4}) and (\ref{9})

\section{Applications}

As we have stated in the introduction, the purpose of determining the
SPD's is for their application to problems where the FF's are insufficient
to describe the physics involved.  We consider in this section two such
problems as illustrations of the usefulness of the SPD's.  The first is when a
hard parton is produced in the environment of thermal partons, as in
heavy-ion collisions.  The second is the determination of two-pion
distribution in a jet.

Let us suppose that a $u$ quark is produced at $k_T = 10$ GeV/c in a
background of thermal partons whose invariant $k_T$ distribution is
\begin{eqnarray}
{\cal T}(k_T) = k_T {dN \over dk_T} = C k_Te^{-k_{T}/T} \quad .
\label{15}
\end{eqnarray}
Let the parameters $C$ and $T$ be chosen to correspond to a typical
situation in Au+Au collisions at $\sqrt{s} = 200$ GeV \cite{hy2}
\begin{eqnarray}
C = 23.2\ {\rm GeV}^{-1},  \qquad  T = 0.317\ {\rm GeV}.
\label{16}
\end{eqnarray}
The high-$k_T$ $u$ quark generates a shower of partons with various
flavors.  Consider specifically $u$ and $\bar{d}$ in that shower.  The
valence quark distribution is given by $K_{NS}(x_1)$, while the
$\bar{d}$ sea-quark distribution (including the ones converted from
the gluons) is given by $L(x_1)$.  In Fig. 3 we plot $dN/k_T dk_T$ for
(a) $u$ quark (valence and sea) in solid line, (b) $\bar{d}$ sea
antiquark in dash-dot line, and (c) $\bar{d}$ thermal antiquark in
dashed line.  They correspond to $k^{-2}_T \times $ (invariant
distributions $K = K_{NS} + L$, $L$, and $\cal{T}$, respectively), in
which $K(x_1)$ and $L(x_1)$ are evaluated at $k_T = x_1 k^{\rm max}_T$,
with $k^{\rm max}_T = 10$\ GeV/c.  Note that the thermal distribution is higher
than the shower parton distributions for $k_T < 1$ GeV/c.  That makes
a crucial difference in the recombination of those partons.  Such a
thermal distribution is absent in $pp$ collisions, whose soft partons
are at least two orders of magnitude lower.  In $e^+e^-$ annihilation
there are, of course, no soft partons at all.

\begin{figure}[tbph]
\includegraphics[width=0.45\textwidth]{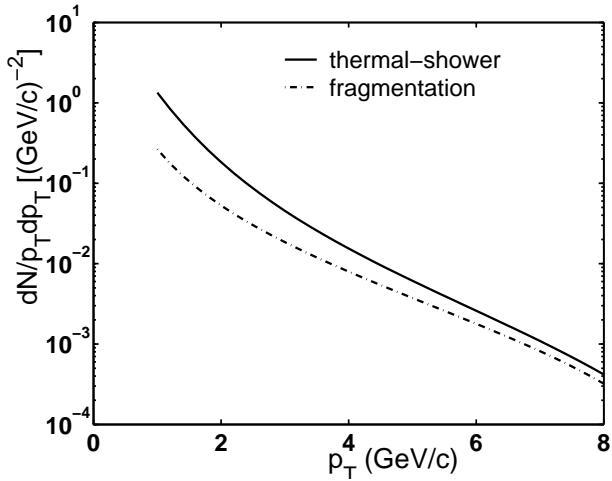}
\caption{
Distributions of $\pi^+$ in $p_T$ arising from
thermal-shower recombination (solid line) and shower-shower
recombination, i.e. fragmentation (dash-dot line).}
\end{figure}

We now calculate the production of $\pi^+$ from the assemblage of
$u$ and $\bar{d}$ partons.  The thermal-shower ($\cal {TS}$)
recombination gives rise to
\begin{eqnarray}
{dN^{{\cal TS}}_{\pi^+}  \over  p_T dp_T} = {1  \over  p^3_T}
\int^{p_T}_0 dk_T K (k_T/k^{\rm max}_T){\cal T}(p_T-k_T),
\label{17}
\end{eqnarray}
where Eq.\ (\ref{10}) has been used in an equation such as Eq.\
(\ref{1}) for $xdN_{\pi}/dx$, but expressed for $dN_{\pi}/p_Tdp_T$.
Using Eqs.\ (\ref{3}), (\ref{15}) and the parametrizations given in Table
II, the integral in Eq.\ (\ref{17}) can readily be evaluated.  The result is
shown by the solid line in Fig.\ 4.  It is  to be compared with the $p_T$
distribution from the fragmentation of the $u$ quark to $\pi^+$, which
is
\begin{eqnarray}
{dN^{{\cal SS}}_{\pi^+}\over  p_T dp_T}={1\over p^3_T}
\int^{p_T}_0dk_T\left\{K\left({k_T\over k^{\rm max}_T}\right),
L\left(p_T-k_T\over k^{\rm max}_T-k_T\right)
\right\},
\label{18}
\end{eqnarray}
Since this is just retracing the path in which we obtained $K$ and $L$
from the $D$ function in the first place, Eq.\ (\ref{18}) can more
directly be identified with
\begin{eqnarray}
{dN^{frag}_{\pi^+}\over p_Tdp_T}=\left(p_Tk^{\rm max}_T\right)^{-1}
\left[D^{\pi}_V\left({p_T\over k^{\rm max}_T}\right)+
D^{\pi}_S\left(p_T\over k^{\rm max}_T\right)
\right] .
\label{19}
\end{eqnarray}
The result is shown by the dash-dot line in Fig.\ 4.  Evidently, the
contribution from the thermal-shower recombination is much more
important than that from fragmentation in the range of $p_T$ shown.
Despite the fact that ${\cal T}(k_T)$
is lower than $L(k_T)$ for $k_T > 1.5$ GeV/c, its dominance at
$k_T < 1.5$ GeV/c is enough to result in the ${\cal TS}$ recombination
to dominate over the ${\cal SS}$ recombination for all $p_T<8$ GeV/c.
This example demonstrates the necessity of knowing the SPD's in a
jet, since $K(x_1)$ is used in Eq.\ (\ref{17}). If $\cal SS$
recombination is the only important contribution as in $pp$
collisions, then fragmentation as in Eq.\ (\ref{19}) is all that is
needed, and the search for SPD's plays no crucial role. In realistic
problems the
hard-parton momentum $k^{\rm max}_T$ has to be integrated over the
weight of the jet cross section.  However, for our illustrative purpose
here, that is beside the point.

Our next example is the study of the dihadron distribution in a jet.  We
need only carry out the investigation here for a jet in vacuum, since the
replacement of  a shower parton by a thermal parton for a jet in a
medium is trivial, having seen how that is done in the replacement of
Eq.\ (\ref{18}) by (\ref{17}) in the case of the single-particle distribution.
Consider the joint distribution of two $\pi^+$ in a jet initiated by a hard
$u$ quark.  As we shall work in the momentum fraction variables, the
value of the momentum of the initiating $u$ quark is irrelevant, except
that it should be high.  Let $X_1$ and $X_2$ denote the momentum
fractions of the two $\pi^+$, and $x_i$ denotes that of the $i$th parton,
$i = 1, \cdots, 4$.  Then, since only one $u$ quark can be valence, the
other three quarks being in the sea, we have one $K$, three $L$, and
two $R$ functions.  Combinatorial complications arise when we impose
the condition that $\sum_i x_i < 1$ for $i = 1,2,3,4$.  There are two
methods to keep the accounting of the different orderings of the four
$x_i$.

\begin{widetext}
{\it Method 1.}

Let one ordering be
\begin{eqnarray}
SPD (x_1x_2x_3x_4) = K(x_1)L\left({ x_2 \over 1-x_1 } \right)L\left({
x_3 \over 1-x_1-x_2 } \right)L\left({ x_4 \over 1-x_1-x_2-x_3 } \right) .
\label{20}
\end{eqnarray}
There are 4! ways to rearrange the four $x_i$ in all orders.  However,
they are to be convoluted with $R_{\pi}(x_1, x_2, X_1)$, which is
symmetric in $x_1 \leftrightarrow x_2$, and similarly with
$R_{\pi}(x_3, x_4, X_2)$.  Thus there are $4!/2!2!$ independent terms.
Since $K$ can appear at any one of the four positions in Eq.\ (\ref{20}),
we have altogether 24 terms.  Thus we have
\begin{eqnarray}
X_1X_2{dN_{\pi^+ \pi^+}  \over dX_1 dX_2 } = \int
\left(\prod^4_{i = 1} {dx_i  \over  x_i}\right)\
  \left[ { 1 \over
24}\sum_P SPD (x_1x_2x_3x_4)\right]
R_{\pi}(x_1,x_2,X_1) R_{\pi}(x_3,x_4,X_2) ,
\label{21}
\end{eqnarray}
where $\sum_P$ symbolizes the permutation of all $x_i$ and summing
over all four positions of $K$, but eliminating redundant terms that
are symmetric under the interchanges of $x_1 \leftrightarrow x_2$ and
$x_3 \leftrightarrow x_4$.

{\it Method 2.}

Let us fix the ordering in Eq.\ (\ref{20}) but permute the contributing
$x_i$ to $X_1$ and $X_2$.  There are six arrangements of $x_i$ and $x_j$ in
$R_{\pi}(x_i, x_j, X_1)R_{\pi}(x_{i'}, x_{j'}, X_2)$, while counting
in $x_{i'}$ and $x_{j'}$ is unnecessary. Let us denote the
summation over them by $\sum_Q$.  Thus we have
\begin{eqnarray}
X_1X_2{dN_{\pi^+ \pi^+}  \over dX_1 dX_2 } = \int
\left(\prod^4_{i = 1} {dx_i  \over  x_i}\right)
   \left[ { 1 \over
4}\sum_K SPD (x_1x_2x_3x_4)\right]
\left[ { 1 \over
6}\sum_Q R_{\pi}(x_i,x_j,X_1) R_{\pi}(x_{i'},x_{j'},X_2)\right]
\label{22}
\end{eqnarray}
where $\sum_K$ denotes summing over the four positions of $K$.
Equation (\ref{22}) is equivalent to (\ref{21}).

It should be noted that not all terms in these equations can be
expressed in the form factorizable FF's.  One example that can is
\begin{eqnarray}
&&{\int} \left(\prod^4_{i = 1} {dx_i  \over  x_i}\right) {1  \over
2}\left[K(x_1)L\left({x_2  \over 1-x_1}\right) + L(x_1) K\left({x_2
\over 1-x_1}\right)\right] R_{\pi}(x_1,x_2,X_1)\nonumber\\
&&\times
L\left({ x_3
\over 1-x_1-x_2 } \right)L\left({ x_4 \over 1-x_1-x_2-x_3 }
\right)R_{\pi}(x_3,x_4,X_2)\nonumber\\
&&= D^{\pi^+}_u(X_1)
D^{\pi^+}_S\left(X_2/(1-X_1)\right) .\label{23}
\end{eqnarray}
\end{widetext}
Because of the presence of terms that cannot be written in factorizable
form, the two-particle distribution cannot be adequately represented by
the FF's only.

\begin{figure}[tbph]
\includegraphics[width=0.45\textwidth]{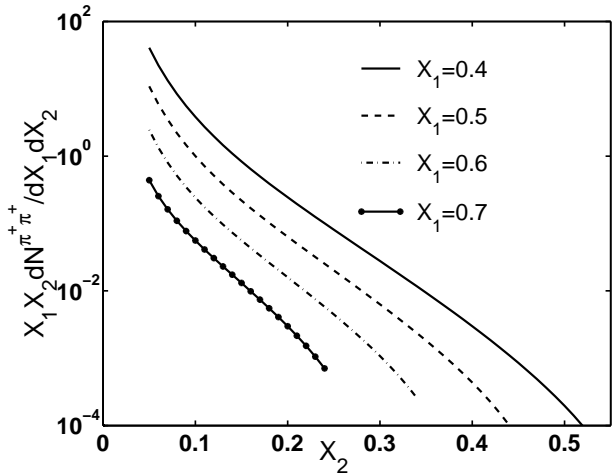}
\caption{
Two $\pi ^+$ correlated distribution in a  $u$-quark
initiated jet.}
\end{figure}

Using the SPD's obtained in the previous section, we get the results
shown in Fig.\ 5, which exhibits the $X_2$ distribution for four fixed
values of $X_1$.  This type of correlation in parton fragmentation has
never been calculated before.  Although the shapes of the $X_2$
distributions look similar in the log scale in Fig.\ 5, there is significant
attenuation as $X_2 \rightarrow 1-X_1$ for each value of $X_1$.  Thus
the effective slope becomes steeper for larger $X_1$.  Recent
experiments at RHIC have begun to measure the distribution of particles
associated with triggers restricted to a small interval. The extension of
our calculation here to such problems will need the input of jet cross
sections for all hard partons in heavy-ion collisions and the participation
of thermal partons in the recombination.  Here we only demonstrate
the utility of the SPD's in the study of dihadron correlation.

\section{Conclusion}

We have described the fragmentation process in the framework of
recombination.  The shower parton distributions obtained are shown to
be useful in problems where the knowledge of the fragmentation
functions alone is not sufficient to provide answers to questions
concerning the interaction between a jet and its surrounding medium or
between particles within a jet.  Such questions arise mainly in nuclear
collisions at high energies.

In our view the basic hadronization process is recombination, even for
fragmentation in vacuum.  Since the recombination process can only be
formulated in the framework of a model, the shower parton
distributions obtained are indeed model dependent.  That is a price that
must be paid for the study of hadrons produced at intermediate $p_T$
where the interaction between soft and semi-hard partons cannot be
ignored, and where perturbative QCD is not reliable.  Once
recombination is adopted for treating hadronization in that $p_T$
range, the extension to higher
$p_T$ can remain in the recombination framework, since the
fragmentation process is recovered by the recombination of two shower
partons.  For hadron production in heavy-ion collisions at super high
energies, such as at LHC, then the high density of hard partons
produced will require the consideration of recombination of hard
partons from overlapping jets.  Thus it is sensible to remain in  the
recombination mode for all $p_T$.

We have shown in this paper how the SPD's can be determined from the
FF's.  Although we have determined the SPD's at only one value of $Q^2$
for the FF's, it is clear that the same procedure can be followed for other
value of $Q^2$.  The formal description of how the $Q^2$ dependences
of the FF's can be transferred to the $Q^2$ dependences of the SPD's is a
problem that is worth dedicated attention.  While the numerical
accuracy of the SPD's obtained here can still be improved, especially at
lower $Q^2$, for the purpose of phenomenological applications the
availability of the parametrizations given in Table II is far more
important than not taking into account at all the shower partons and
their interactions with the medium in the environment.  The $Q^2$
evolution of the SPD's may have to undergo a long process of
investigatory evolution of its own just as what has happened to the FF's.
That can proceed in parallel to the rich phenomenology that can now
be pursued in the application of the role of shower partons to
heavy-ion collisions.
\begin{center}
\section*{Acknowledgment}
\end{center}
We are grateful to S.\ Kretzer for a helpful communication.
This work was supported, in
part,  by the U.\ S.\ Department of Energy under Grant No.
DE-FG03-96ER40972  and by the Ministry of Education of China under
Grant No. 03113.

\end{document}